# Acoustic microstreaming and shear stress produced by the interaction of an oscillating gas bubble with a viscoelastic particle

Alexander A. Doinikov[1], Cyril Mauger[1], Philippe Blanc-Benon[1] and Claude Inserra[2,†]

[1]INSA de Lyon, CNRS, École Centrale de Lyon, Université Claude Bernard Lyon 1, LMFA, UMR 5509, 69621 Villeurbanne, France

[2]Univ Lyon, Université Claude Bernard Lyon 1, Centre Léon Bérard, INSERM, UMR 1032, LabTAU, F-69003 Lyon, France

†Email address for correspondence: claude.inserra@inserm.fr

An analytical theory is developed that describes acoustic microstreaming produced by the interaction of an oscillating gas bubble with a viscoelastic particle. The bubble is assumed to undergo axisymmetric oscillation modes, which can include radial oscillation, translation and shape modes. The oscillations of the particle are excited by the oscillations of the bubble. No restrictions are imposed on the ratio of the bubble and the particle radii to the viscous penetration depth and the separation distance, as well as on the ratio of the viscous penetration depth to the separation distance. Capabilities of the developed theory are illustrated by computational examples. The shear stress produced by the acoustic microstreaming on the particle surface is calculated. It is shown that this stress is much higher than the stress predicted by Nyborg's formula [J. Acoust. Soc. Am. **30**, 329–339 (1958)], which is commonly used to evaluate the time-averaged shear stress produced by a bubble on a rigid wall.

# 1. Introduction

Studies on the acoustic interaction between an oscillating gas bubble and a solid particle are motivated by biomedical and microfluidic applications such as hemolysis, sonoporation and cell manipulation in microchannel devices (Tachibana *et al*. 1999; Wu *et al*. 2002; Marmottant & Hilgenfeldt 2003; Deng *et al*. 2004; Mehier-Humbert *et al*. 2005; van Wamel *et al*. 2006; Wu & Nyborg 2008; Zinin & Allen III 2009). This interaction is characterized by two main effects: radiation interaction force, called also secondary acoustic radiation force (Saeidi *et al*. 2020), and acoustic microstreaming (Nyborg 1965).

The acoustic microstreaming that occurs between a bubble and a particle produces shear stress on the particle surface. It is assumed that this effect, depending on its strength, can cause cell disruption or sonoporation (temporary "opening" of the plasma membrane of a cell for the incorporation of foreign macromolecules into the cell without serious consequences for the cell viability) (Rooney 1972; Lewin & Bjørnø 1982; Wu 2002, 2007; VanBavel 2007; Doinikov & Bouakaz 2010).

To evaluate theoretically the time-averaged shear stress produced by a bubble on a particle (cell), the interaction between the bubble and the particle is commonly modeled as the interaction of a bubble with a plane rigid wall. Nyborg (1958) has derived an analytical formula for this model. However, his formula is not based on an exact mathematical solution of the problem. It is based on an approximate solution that assumes the viscous boundary layer thickness to be small compared to the scale of the distribution of the oscillatory fluid velocity. This assumption implies that the acoustic streaming exists only in the near-boundary region, i.e., within a thin sheet of fluid at the solid boundary. The fluid motion outside this region is assumed potential and the irrotational part of the oscillatory fluid velocity is only taken into account.

In the present paper, we develop an analytical theory that describes acoustic microstreaming produced by the interaction of an oscillating gas bubble with a viscoelastic particle at an arbitrary separation distance between them. The bubble is assumed to undergo axisymmetric oscillation modes, which can include radial oscillation, translation and shape modes. The oscillations of the particle are excited by the oscillations of the bubble. No restrictions are imposed on the ratio of the bubble and the particle radii to the viscous penetration depth and the separation distance, as well as on the ratio of the viscous penetration depth to the separation distance. The developed theory is used to calculate the shear stress produced by the acoustic microstreaming on the particle surface.

## 2. Theory

We consider a gas bubble and a viscoelastic particle surrounded by a viscous incompressible liquid. Both the bubble and the particle are assumed to be spherical at rest. We use two spherical coordinate systems, $(r_1, \theta_1, \varepsilon_1)$ and $(r_2, \theta_2, \varepsilon_2)$, which are originated at the equilibrium centres of the bubble and the particle, respectively; see figure 1. The direction $\theta_1 = \theta_2 = 0$ corresponds to the $z$ axis. The distance between the equilibrium centres of the bubble and the particle is denoted by $d$. Our calculation is based on the theory developed by Doinikov *et al.* (2022) for acoustic microstreaming generated by two interacting oscillating bubbles.

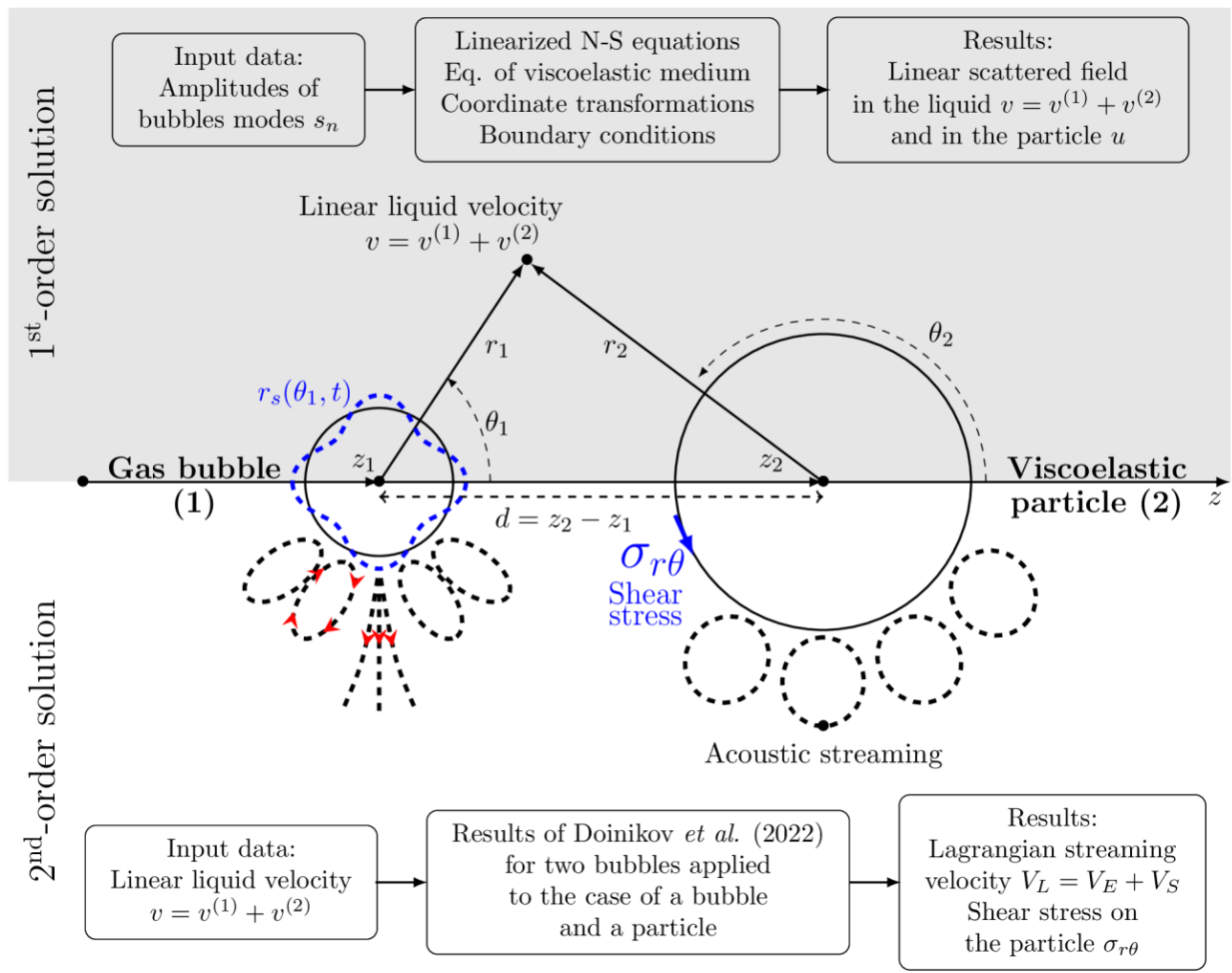

Figure 1. Coordinate systems used in calculations and calculation flowcharts for the first-order solution (top) and the second-order solution (bottom). $r_s$ is the radial coordinate of the disturbed bubble surface, $v^{(1)}$ and $v^{(2)}$ are the first-order liquid velocities generated by the oscillations of the bubble and the particle, respectively, $u$ is the displacement vector inside the particle, $V_E$ and $V_S$ are, respectively, the Eulerian streaming velocity and the Stokes drift velocity in the liquid.

We assume that in the general case, the bubble undergoes $N$ axisymmetric oscillation modes with mode numbers $M_1$, $M_2$,..., $M_N$. Then the disturbed surface of the bubble can be represented by

$$r_s = R_{10} + e^{-i\omega t} \sum_{n=M_1}^{M_N} s_n P_n(\mu_1), \qquad (2.1)$$

where $r_s$ is the radial coordinate of the disturbed bubble surface, $R_{10}$ is the equilibrium radius of the bubble, $\omega$ is the angular oscillation frequency, $P_n$ is the Legendre polynomial of degree $n$, $\mu_1 = \cos\theta_1$ and $s_n$ is the complex amplitude of the $n$th mode of the bubble. The values of $s_n$ can be taken from experimental measurements or, in some cases, they can be calculated analytically in terms of the amplitude of the driving acoustic pressure. We assume that the oscillations of the bubble are excited by the driving acoustic pressure, while the oscillations of the particle are excited by the oscillations of the bubble.

The process of the derivation of the second-order mean flow (acoustic streaming) around the bubble-particle system, as well as the calculation of the time-averaged shear stress on the particle surface, is summarized in figure 1 using a flowchart representation. The supposedly known amplitudes of the bubble oscillations are used as input data when deriving the first-order (linear) scattered field in the liquid (the top flowchart in figure 1). In parallel, the first-order scattered field inside the particle is derived using the equation of motion of a viscoelastic medium. The unknown constants of the first-order solution are found using appropriate boundary conditions at the interfaces of the bubble and the particle. At the next stage of the calculation (the bottom flowchart in figure 1), the time-averaged second-order liquid velocity is derived. To this end, the results obtained by Doinikov *et al.* (2022) for the case of two interacting nonspherically-oscillating bubbles are applied to the bubble-particle system under study, using the first-order liquid velocity field obtained here at the first stage of the calculation. Doing so, the right bubble in the derivation of Doinikov *et al*. (2022) is replaced by the viscoelastic particle. This approach provides the Lagrangian streaming velocity field around the bubble-particle system, which then allows one to calculate the time-averaged shear stress on the particle surface.

2.1. *First-order scattered field in the liquid*

We begin with the calculation of the first-order scattered acoustic field in the liquid. To this end, we use the linearized equations of a viscous incompressible liquid, which are given by (Landau & Lifshitz 1987)

$$\nabla \cdot \mathbf{v} = 0, \tag{2.2}$$

$$\frac{\partial \mathbf{v}}{\partial t} = -\frac{1}{\rho}\nabla p + \nu \Delta \mathbf{v}, \tag{2.3}$$

where $\mathbf{v}$ is the first-order liquid velocity, $p$ is the first-order liquid pressure, $\rho$ is the liquid density, $\nu = \eta/\rho$ is the kinematic liquid viscosity and $\eta$ is the dynamic liquid viscosity.

Since we have two sources of scattering, the bubble and the particle, the velocity $\mathbf{v}$ can be written as

$$\mathbf{v} = \mathbf{v}^{(1)} + \mathbf{v}^{(2)}, \tag{2.4}$$

where $\mathbf{v}^{(1)}$ and $\mathbf{v}^{(2)}$ are the first-order liquid velocities generated by the oscillations of the bubble and the particle, respectively.

The velocity $\mathbf{v}^{(j)}$ is represented by

$$\mathbf{v}^{(j)} = \nabla \varphi^{(j)} + \nabla \times \boldsymbol{\psi}^{(j)}, \tag{2.5}$$

where $\varphi^{(j)}$ and $\boldsymbol{\psi}^{(j)}$ are the scalar and the vector velocity potentials, respectively.

It is shown by Doinikov *et al*. (2022) that in the case of axial symmetry, $\varphi^{(j)}$ and $\boldsymbol{\psi}^{(j)}$ are given by

$$\varphi^{(j)}(r_j,\theta_j,t) = e^{-i\omega t}\sum_{n=0}^{\infty} a_n^{(j)}\left(\frac{R_{j0}}{r_j}\right)^{n+1} P_n(\mu_j), \tag{2.6}$$

$$\boldsymbol{\psi}^{(j)}(r_j,\theta_j,t) = e^{-i\omega t}\mathbf{e}_{\varepsilon j}\sum_{n=1}^{\infty} b_n^{(j)} h_n^{(1)}(k_\nu r_j) P_n^1(\mu_j), \tag{2.7}$$

where $h_n^{(1)}$ is the spherical Hankel function of the first kind, $k_\nu = (1+i)/\delta$ is the viscous wavenumber, $\delta = \sqrt{2\nu/\omega}$ is the viscous penetration depth, $P_n^1$ is the associated Legendre polynomial of the first order and degree $n$, $\mu_j = \cos\theta_j$, $R_{20}$ is the equilibrium radius of the particle and $\boldsymbol{e}_{\varepsilon j}$ is the unit azimuth vector of the $j$th coordinate system. The axial symmetry allows us to set $\varepsilon_1 = \varepsilon_2$ and $\boldsymbol{e}_{\varepsilon 1} = \boldsymbol{e}_{\varepsilon 2}$. The constants $a_n^{(j)}$ and $b_n^{(j)}$, called linear scattering coefficients, are calculated by boundary conditions at the surfaces of the bubble and the particle; see below.

Substitution of (2.6) and (2.7) into (2.5) gives the radial and tangential components of $\boldsymbol{v}^{(j)}$,

$$v_r^{(j)}(r_j,\theta_j,t) = -\frac{e^{-i\omega t}}{r_j}\sum_{n=0}^{\infty}(n+1)\left[a_n^{(j)}\left(\frac{R_{j0}}{r_j}\right)^{n+1} + nb_n^{(j)}h_n^{(1)}(k_v r_j)\right]P_n(\mu_j), \qquad (2.8)$$

$$v_\theta^{(j)}(r_j,\theta_j,t) = \frac{e^{-i\omega t}}{r_j}\sum_{n=1}^{\infty}\left\{a_n^{(j)}\left(\frac{R_{j0}}{r_j}\right)^{n+1} - b_n^{(j)}\left[h_n^{(1)}(k_v r_j) + k_v r_j h_n^{(1)\prime}(k_v r_j)\right]\right\}P_n^1(\mu_j), \qquad (2.9)$$

where the prime denotes the derivative with respect to the argument in brackets.

The first-order liquid pressure $p$ can also be written as

$$p = p^{(1)} + p^{(2)}. \qquad (2.10)$$

Substituting (2.5) into (2.3) and taking into account that $\psi^{(j)}$ obeys the equation $(\Delta + k_v^2)\psi^{(j)} = 0$ (Doinikov et al. 2019), one obtains

$$p^{(j)} = i\omega\rho\varphi^{(j)}. \qquad (2.11)$$

2.2. *First-order scattered field inside the particle*

We assume that the motion of the viscoelastic medium inside the particle obeys the following equation (Landau & Lifshitz 1970):

$$\rho_p\frac{\partial^2 \boldsymbol{u}}{\partial t^2} = \mu_p\nabla^2\boldsymbol{u} + (\lambda_p + \mu_p)\nabla(\nabla\cdot\boldsymbol{u}) + \eta_p\nabla^2\frac{\partial \boldsymbol{u}}{\partial t} + \left(\zeta_p + \frac{1}{3}\eta_p\right)\nabla\left(\nabla\cdot\frac{\partial \boldsymbol{u}}{\partial t}\right), \qquad (2.12)$$

where $\boldsymbol{u}$ is the displacement vector, $\rho_p$ is the particle density, $\lambda_p = E\sigma/[(1-2\sigma)(1+\sigma)]$ and $\mu_p = E/[2(1+\sigma)]$ are the Lamé coefficients, $E$ is Young's modulus, $\sigma$ is Poisson's ratio, $\zeta_p$ is the bulk viscosity and $\eta_p$ is the shear viscosity. The time dependence of $\boldsymbol{u}$ is assumed to be $\exp(-i\omega t)$.

A solution to (2.12) is sought as

$$\boldsymbol{u} = \nabla\varphi_p + \nabla\times\boldsymbol{\psi}_p. \qquad (2.13)$$

Substitution of (2.13) into (2.12) yields

$$\nabla^2 \varphi_p + k_l^2 \varphi_p = 0, \tag{2.14}$$

$$\nabla^2 \boldsymbol{\psi}_p + k_t^2 \boldsymbol{\psi}_p = 0, \tag{2.15}$$

where $k_l$ and $k_t$ are the longitudinal and the transverse wavenumbers, respectively, which are calculated by (Landau & Lifshitz 1970)

$$k_l = \omega \sqrt{\frac{\rho_p}{\lambda_p + 2\mu_p - i\omega(\zeta_p + 4\eta_p/3)}}, \tag{2.16}$$

$$k_t = \omega \sqrt{\frac{\rho_p}{\mu_p - i\omega\eta_p}}. \tag{2.17}$$

Axisymmetric solutions to (2.14) and (2.15) are given by

$$\varphi_p(r_2, \theta_2, t) = e^{-i\omega t} \sum_{n=0}^{\infty} \hat{a}_n j_n(k_l r_2) P_n(\mu_2), \tag{2.18}$$

$$\boldsymbol{\psi}_p(r_2, \theta_2, t) = \boldsymbol{e}_{\varepsilon 2} e^{-i\omega t} \sum_{n=1}^{\infty} \hat{b}_n j_n(k_t r_2) P_n^1(\mu_2), \tag{2.19}$$

where $j_n$ is the spherical Bessel function of order $n$ and $\hat{a}_n$, $\hat{b}_n$ are constants to be determined by boundary conditions.

Substitution of (2.18) and (2.19) into (2.13) gives the radial and tangential components of $\boldsymbol{u}$,

$$u_r(r_2, \theta_2, t) = e^{-i\omega t} \sum_{n=0}^{\infty} \left[ k_l \hat{a}_n j_n'(k_l r_2) - n(n+1) \hat{b}_n \frac{j_n(k_t r_2)}{r_2} \right] P_n(\mu_2), \tag{2.20}$$

$$u_\theta(r_2, \theta_2, t) = \frac{e^{-i\omega t}}{r_2} \sum_{n=1}^{\infty} \left\{ \hat{a}_n j_n(k_l r_2) - \hat{b}_n \left[ j_n(k_t r_2) + k_t r_2 j_n'(k_t r_2) \right] \right\} P_n^1(\mu_2). \tag{2.21}$$

### 2.3. *First-order boundary conditions*

To calculate $a_n^{(j)}$, $b_n^{(j)}$, $\hat{a}_n$ and $\hat{b}_n$, we use the following boundary conditions:

$$v_r = \frac{\partial r_s}{\partial t} \quad \text{at} \quad r_1 = R_{10}, \tag{2.22}$$

$$\sigma_{r\theta} = 0 \quad \text{at} \quad r_1 = R_{10}, \tag{2.23}$$

$$v_r = \frac{\partial u_r}{\partial t} \quad \text{at} \quad r_2 = R_{20}, \tag{2.24}$$

$$v_\theta = \frac{\partial u_\theta}{\partial t} \quad \text{at} \quad r_2 = R_{20}, \tag{2.25}$$

$$\sigma_{rr} = \hat{\sigma}_{rr} \quad \text{at} \quad r_2 = R_{20}, \tag{2.26}$$

$$\sigma_{r\theta} = \hat{\sigma}_{r\theta} \quad \text{at} \quad r_2 = R_{20}, \tag{2.27}$$

where $\sigma_{rr}$ and $\sigma_{r\theta}$ are the first-order normal and tangential stresses in the liquid and $\hat{\sigma}_{rr}$ and $\hat{\sigma}_{r\theta}$ are the normal and tangential stresses in the particle.

The physical meaning of (2.22) – (2.27) is as follows. Equation (2.22) requires that the normal component of the liquid velocity at the bubble surface be equal to the normal component of the surface velocity of the bubble. Equation (2.23) requires that the liquid tangential stress vanish at the bubble surface because the gas viscosity is much lower than the liquid viscosity. Equations (2.24) – (2.27) require that the velocity and the stress be continuous across the surface of the particle. Note that we do not use the boundary condition for the normal stress at the bubble surface. The point is that this condition is commonly used to calculate the amplitudes of the bubble oscillation modes in terms of the amplitude of the imposed acoustic pressure. We assume, however, that the amplitudes of the bubble oscillation modes are known (measured experimentally). This approach allows one to cover the case that shape modes are excited parametrically. Therefore, the boundary condition for the normal stress at the bubble surface is redundant in our derivation.

The stress components are calculated by (Landau & Lifshitz 1970, 1987)

$$\sigma_{rr} = -p + 2\eta \frac{\partial v_r}{\partial r_j}, \tag{2.28}$$

$$\sigma_{r\theta} = \eta \left( \frac{1}{r_j} \frac{\partial v_r}{\partial \theta_j} + \frac{\partial v_\theta}{\partial r_j} - \frac{v_\theta}{r_j} \right), \tag{2.29}$$

$$\hat{\sigma}_{rr} = \lambda_p \nabla \cdot \boldsymbol{u} + 2\mu_p \frac{\partial u_r}{\partial r_2} + 2\eta_p \frac{\partial^2 u_r}{\partial t \partial r_2} + \left( \zeta_p - \frac{2}{3}\eta_p \right) \nabla \cdot \frac{\partial \boldsymbol{u}}{\partial t}, \tag{2.30}$$

$$\hat{\sigma}_{r\theta} = \mu_p \left( \frac{1}{r_2} \frac{\partial u_r}{\partial \theta_2} + \frac{\partial u_\theta}{\partial r_2} - \frac{u_\theta}{r_2} \right) + \eta_p \frac{\partial}{\partial t} \left( \frac{1}{r_2} \frac{\partial u_r}{\partial \theta_2} + \frac{\partial u_\theta}{\partial r_2} - \frac{u_\theta}{r_2} \right). \tag{2.31}$$

### 2.4. Calculation of the linear scattering coefficients

In order to apply (2.22) – (2.27), we need expressions for $\varphi^{(1)}$, $v_r^{(1)}$ and $v_\theta^{(1)}$ in the coordinates $(r_2, \theta_2, \varepsilon_2)$ and expressions for $\varphi^{(2)}$, $v_r^{(2)}$ and $v_\theta^{(2)}$ in the coordinates $(r_1, \theta_1, \varepsilon_1)$. These expressions are given by (A5), (A6), (A11), (A13), (A15) and (A16) of (Doinikov *et al.* 2022),

$$\varphi^{(1)}(r_2, \theta_2, t) = e^{-i\omega t} \sum_{n,m=0}^{\infty} a_n^{(1)} \xi_1^{n+1} (-1)^m C_{nm} \left(\frac{r_2}{d}\right)^m P_m(\mu_2), \tag{2.32}$$

$$v_r^{(1)}(r_2, \theta_2, t) = e^{-i\omega t} \sum_{n,m=0}^{\infty} P_m(\mu_2) \left\{ \frac{(-1)^m m}{d} C_{nm} \xi_1^{n+1} a_n^{(1)} \left(\frac{r_2}{d}\right)^{m-1} \right.$$

$$\left. - \frac{j_m(k_v r_2)}{r_2} \frac{\sqrt{n(n+1)}}{(2n+1)i^n} (2m+1)i^m \sqrt{m(m+1)} b_n^{(1)} \sum_{l=0}^{\infty} (-1)^l i^l (2l+1) C_{m0l0}^{n0} C_{m1l0}^{n1} h_l^{(1)}(k_v d) \right\}, \tag{2.33}$$

$$v_\theta^{(1)}(r_2, \theta_2, t) = \frac{e^{-i\omega t}}{r_2} \sum_{n,m=0}^{\infty} P_m^1(\mu_2) \left\{ (-1)^m C_{nm} \xi_1^{n+1} a_n^{(1)} \left(\frac{r_2}{d}\right)^m \right.$$

$$\left. - [j_m(k_v r_2) + k_v r_2 j_m'(k_v r_2)] \frac{\sqrt{n(n+1)}}{(2n+1)i^n} \frac{(2m+1)i^m}{\sqrt{m(m+1)}} b_n^{(1)} \sum_{l=0}^{\infty} (-1)^l i^l (2l+1) C_{m0l0}^{n0} C_{m1l0}^{n1} h_l^{(1)}(k_v d) \right\}, \tag{2.34}$$

$$\varphi^{(2)}(r_1, \theta_1, t) = e^{-i\omega t} \sum_{n,m=0}^{\infty} (-1)^n a_n^{(2)} \xi_2^{n+1} C_{nm} \left(\frac{r_1}{d}\right)^m P_m(\mu_1), \tag{2.35}$$

$$v_r^{(2)}(r_1, \theta_1, t) = e^{-i\omega t} \sum_{n,m=0}^{\infty} P_m(\mu_1) \left\{ \frac{(-1)^n m}{d} C_{nm} \xi_2^{n+1} a_n^{(2)} \left(\frac{r_1}{d}\right)^{m-1} \right.$$

$$\left. - \frac{j_m(k_v r_1)}{r_1} \frac{\sqrt{n(n+1)}}{(2n+1)i^n} (2m+1)i^m \sqrt{m(m+1)} b_n^{(2)} \sum_{l=0}^{\infty} i^l (2l+1) C_{m0l0}^{n0} C_{m1l0}^{n1} h_l^{(1)}(k_v d) \right\}, \tag{2.36}$$

$$v_\theta^{(2)}(r_1, \theta_1, t) = \frac{e^{-i\omega t}}{r_1} \sum_{n,m=0}^{\infty} P_m^1(\mu_1) \left\{ (-1)^n C_{nm} \xi_2^{n+1} a_n^{(2)} \left(\frac{r_1}{d}\right)^m \right.$$

$$\left. - [j_m(k_v r_1) + k_v r_1 j_m'(k_v r_1)] \frac{\sqrt{n(n+1)}}{(2n+1)i^n} \frac{(2m+1)i^m}{\sqrt{m(m+1)}} b_n^{(2)} \sum_{l=0}^{\infty} i^l (2l+1) C_{m0l0}^{n0} C_{m1l0}^{n1} h_l^{(1)}(k_v d) \right\}, \tag{2.37}$$

where $\xi_j = R_{j0}/d$, $C_{nm} = (n+m)!/(n!m!)$ and $C_{l_1 m_1 l_2 m_2}^{lm}$ are the Clebsch-Gordan coefficients (Abramowitz & Stegun 1972; Varshalovich *et al.* 1988; Zwillinger 2003).

Substituting (2.1), (2.8) at $j = 1$ and (2.36) into (2.22), one obtains

$$(n+1)a_n^{(1)} + n(n+1)h_n^{(1)}(x_1)b_n^{(1)} - n\xi_1^n \sum_{m=0}^{\infty}(-1)^m C_{nm}\xi_2^{m+1}a_m^{(2)}$$

$$+i^n(2n+1)\sqrt{n(n+1)}\,j_n(x_1)\sum_{m=1}^{\infty}\kappa_{nm}^{(2)}b_m^{(2)} = i\omega R_{10}\sum_{m=M_1}^{M_N}\delta_{nm}s_m,\ n\geq 0, \qquad (2.38)$$

where $x_j = k_v R_{j0}$, $\delta_{nm}$ is the Kronecker delta and $\kappa_{nm}^{(j)}$ is defined by

$$\kappa_{nm}^{(j)} = \frac{\sqrt{m(m+1)}}{(2m+1)i^m}\sum_{l=|n-m|}^{n+m}(-1)^{jl}i^l(2l+1)C_{n0l0}^{m0}C_{n1l0}^{m1}h_l^{(1)}(k_v d). \qquad (2.39)$$

Note that in (2.39), the triangle inequality has been used, which requires that the indices of $C_{l_1 m_1 l_2 m_2}^{lm}$ satisfy the following conditions: $l_1 + l_2 - l \geq 0$, $l_1 - l_2 + l \geq 0$, $-l_1 + l_2 + l \geq 0$ (Zwillinger 2003).

To apply the boundary condition for the tangential stress (2.23), we calculate (2.29) at $j = 1$ using (2.8) and (2.9) at $j = 1$, (2.36) and (2.37). As a result, (2.23) gives

$$2(n+2)a_n^{(1)} + \left[(n^2+n-2)h_n^{(1)}(x_1) + x_1^2 h_n^{(1)\prime\prime}(x_1)\right]b_n^{(1)} + 2(n-1)\xi_1^n\sum_{m=0}^{\infty}(-1)^{m+1}C_{nm}\xi_2^{m+1}a_m^{(2)}$$

$$+\frac{i^n(2n+1)}{\sqrt{n(n+1)}}\left[(n^2+n-2)j_n(x_1) + x_1^2 j_n^{\prime\prime}(x_1)\right]\sum_{m=1}^{\infty}\kappa_{nm}^{(2)}b_m^{(2)} = 0,\ n\geq 1. \qquad (2.40)$$

Substitution of (2.8) at $j = 2$, (2.20) and (2.33) into (2.24) yields

$$(n+1)a_n^{(2)} + n(n+1)h_n^{(1)}(x_2)b_n^{(2)} - (-1)^n n\xi_2^n \sum_{m=0}^{\infty}C_{nm}\xi_1^{m+1}a_m^{(1)}$$

$$+i^n(2n+1)\sqrt{n(n+1)}\,j_n(x_2)\sum_{m=1}^{\infty}\kappa_{nm}^{(1)}b_m^{(1)} - i\omega x_l j_n'(x_l)\widehat{a}_n + i\omega n(n+1)j_n(x_t)\widehat{b}_n = 0,\ n\geq 0, \qquad (2.41)$$

where $x_l = k_l R_{20}$ and $x_t = k_t R_{20}$.

Substitution of (2.9) at $j = 2$, (2.21) and (2.34) into (2.25) yields

$$a_n^{(2)} - \left[h_n^{(1)}(x_2) + x_2 h_n^{(1)\prime}(x_2)\right]b_n^{(2)} + (-1)^n \xi_2^n \sum_{m=0}^{\infty}C_{nm}\xi_1^{m+1}a_m^{(1)}$$

$$-\frac{i^n(2n+1)}{\sqrt{n(n+1)}}\left[j_n(x_2) + x_2 j_n'(x_2)\right]\sum_{m=1}^{\infty}\kappa_{nm}^{(1)}b_m^{(1)} + i\omega j_n(x_l)\widehat{a}_n - i\omega\left[j_n(x_t) + x_t j_n'(x_t)\right]\widehat{b}_n = 0,\ n\geq 1. \qquad (2.42)$$

To apply the boundary condition for the normal stress (2.26), we first calculate (2.28) at $j = 2$ using (2.6) and (2.8) at $j = 2$, (2.11), (2.32) and (2.33). We then calculate (2.30) using (2.20), the fact that $\nabla \cdot \boldsymbol{u} = \nabla^2 \varphi_p = -k_l^2 \varphi_p$ and (2.18). Substituting the resulting expressions for $\sigma_{rr}$ and $\hat{\sigma}_{rr}$ into (2.26), we obtain

$$\left[ i\omega\rho R_{20}^2 - 2(n+1)(n+2)\eta \right] a_n^{(2)} + 2n(n+1)\eta \left[ x_2 h_n^{(1)\prime}(x_2) - h_n^{(1)}(x_2) \right] b_n^{(2)}$$

$$+ (-1)^n \xi_2^n \left[ i\omega\rho R_{20}^2 - 2n(n-1)\eta \right] \sum_{m=0}^{\infty} C_{nm} \xi_1^{m+1} a_m^{(1)}$$

$$+ 2\eta i^n (2n+1) \sqrt{n(n+1)} \left[ x_2 j_n'(x_2) - j_n(x_2) \right] \sum_{m=1}^{\infty} \kappa_{nm}^{(1)} b_m^{(1)}$$

$$+ \left[ \left( i\omega\zeta_p - \frac{2}{3} i\omega\eta_p - \lambda_p \right) j_n(x_l) + 2(\mu_p - i\omega\eta_p) j_n''(x_l) \right] x_l^2 \hat{a}_n$$

$$+ 2n(n+1)(\mu_p - i\omega\eta_p) \left[ j_n(x_t) - x_t j_n'(x_t) \right] \hat{b}_n = 0, \; n \geq 0. \tag{2.43}$$

To apply the boundary condition for the tangential stress (2.27), we first calculate (2.29) at $j = 2$ using (2.8) and (2.9) at $j = 2$, (2.33) and (2.34). We then calculate (2.31) using (2.20) and (2.21). Substituting the resulting expressions for $\sigma_{r\theta}$ and $\hat{\sigma}_{r\theta}$ into (2.27), we obtain

$$2(n+2)a_n^{(2)} + \left[ (n^2 + n - 2)h_n^{(1)}(x_2) + x_2^2 h_n^{(1)\prime\prime}(x_2) \right] b_n^{(2)} - 2(-1)^n (n-1) \xi_2^n \sum_{m=0}^{\infty} C_{nm} \xi_1^{m+1} a_m^{(1)}$$

$$+ \frac{i^n (2n+1)}{\sqrt{n(n+1)}} \left[ (n^2 + n - 2) j_n(x_2) + x_2^2 j_n''(x_2) \right] \sum_{m=1}^{\infty} \kappa_{nm}^{(1)} b_m^{(1)}$$

$$+ \frac{i\omega\eta_p - \mu_p}{\eta} \left\{ 2 \left[ j_n(x_l) - x_l j_n'(x_l) \right] \hat{a}_n + \left[ (n^2 + n - 2) j_n(x_t) + x_t^2 j_n''(x_t) \right] \hat{b}_n \right\} = 0, \; n \geq 1. \tag{2.44}$$

Setting $n = 0$ in (2.38), one obtains

$$a_0^{(1)} = i\omega R_{10} s_0. \tag{2.45}$$

This equation shows that $a_0^{(1)}$ is determined by the amplitude of the bubble radial mode. $a_0^{(1)} \neq 0$ only if $s_0 \neq 0$. Equation (2.45) means that $a_0^{(1)}$ can be considered as a known quantity when the other scattering coefficients are calculated.

Setting $n = 0$ in (2.41) and (2.43), one obtains

$$a_0^{(2)} + i\omega x_l j_1(x_l) \hat{a}_0 = 0, \tag{2.46}$$

$$(i\omega\rho R_{20}^2 - 4\eta) a_0^{(2)} + i\omega\rho R_{20}^2 \sum_{m=0}^{\infty} \xi_1^{m+1} a_m^{(1)}$$

$$+x_l^2\left[\left(i\omega\zeta_p - \frac{2}{3}i\omega\eta_p - \lambda_p\right)j_0(x_l) + 2(\mu_p - i\omega\eta_p)j_0''(x_l)\right]\hat{a}_0 = 0. \tag{2.47}$$

It follows from these equations that

$$a_0^{(2)} = c_1\left(\xi_1 a_0^{(1)} + \sum_{n=1}^{\infty}\xi_1^{n+1} a_n^{(1)}\right), \tag{2.48}$$

$$\hat{a}_0 = c_2 a_0^{(2)}, \tag{2.49}$$

where

$$c_1 = \left\{\frac{x_l}{\rho\omega^2 R_{20}^2 j_1(x_l)}\left[\left(\lambda_p - i\omega\zeta_p + \frac{2i\omega\eta_p}{3}\right)j_0(x_l) + 2(\mu_p - i\omega\eta_p)j_1'(x_l)\right] + \frac{4\eta}{i\omega\rho R_{20}^2} - 1\right\}^{-1}, \tag{2.50}$$

$$c_2 = \frac{i}{\omega x_l j_1(x_l)}. \tag{2.51}$$

Equations (2.48) and (2.49) show that $a_0^{(2)}$ and $\hat{a}_0$ are expressed in terms of $a_0^{(1)}$ and $a_n^{(1)}$ with $n \geq 1$. This means that $a_0^{(2)}$ and $\hat{a}_0$ can be eliminated from the calculation of the coefficients with $n \geq 1$.

For $n \geq 1$, combining (2.38) and (2.40) – (2.44) with (2.45), (2.48) and (2.49), one obtains

$$\sum_{m=1}^{\infty}\left[\delta_{mn}(n+1) - n\xi_1^{n+m+1}\xi_2 c_1\right]a_m^{(1)} + \delta_{mn}n(n+1)h_n^{(1)}(x_1)b_m^{(1)}$$

$$+(-1)^{m+1}nC_{nm}\xi_1^n\xi_2^{m+1}a_m^{(2)} + i^n(2n+1)\sqrt{n(n+1)}\,j_n(x_1)\kappa_{nm}^{(2)}b_m^{(2)}$$

$$= i\omega R_{10}\left(n\xi_1^{n+1}\xi_2 c_1 s_0 + s_n\right), \tag{2.52}$$

$$\sum_{m=1}^{\infty}2\left[\delta_{mn}(n+2) - (n-1)\xi_1^{n+m+1}\xi_2 c_1\right]a_m^{(1)} + \delta_{mn}\left[(n^2+n-2)h_n^{(1)}(x_1) + x_1^2 h_n^{(1)''}(x_1)\right]b_m^{(1)}$$

$$+2(-1)^{m+1}(n-1)C_{nm}\xi_1^n\xi_2^{m+1}a_m^{(2)} + \frac{i^n(2n+1)}{\sqrt{n(n+1)}}\left[(n^2+n-2)j_n(x_1) + x_1^2 j_n''(x_1)\right]\kappa_{nm}^{(2)}b_m^{(2)}$$

$$= 2i(n-1)\omega R_{10}\xi_1^{n+1}\xi_2 c_1 s_0, \tag{2.53}$$

$$\sum_{m=1}^{\infty}(-1)^{n+1}nC_{nm}\xi_1^{m+1}\xi_2^n a_m^{(1)} + i^n(2n+1)\sqrt{n(n+1)}\,j_n(x_2)\kappa_{nm}^{(1)}b_m^{(1)}$$

$$+(n+1)a_n^{(2)} + n(n+1)h_n^{(1)}(x_2)b_n^{(2)} - i\omega x_l j_n'(x_l)\hat{a}_n + i\omega n(n+1)j_n(x_t)\hat{b}_n$$

$$= i(-1)^n n\omega R_{10}\xi_1\xi_2^n s_0, \tag{2.54}$$

$$\sum_{m=1}^{\infty}(-1)^n C_{nm}\xi_1^{m+1}\xi_2^n a_m^{(1)} - \frac{i^n(2n+1)}{\sqrt{n(n+1)}}\left[j_n(x_2) + x_2 j_n'(x_2)\right]\kappa_{nm}^{(1)}b_m^{(1)}$$
$$+a_n^{(2)} - \left[h_n^{(1)}(x_2) + x_2 h_n^{(1)\prime}(x_2)\right]b_n^{(2)} + i\omega j_n(x_l)\widehat{a}_n - i\omega\left[j_n(x_t) + x_t j_n'(x_t)\right]\widehat{b}_n$$
$$= i(-1)^{n+1}\omega R_{10}\xi_1\xi_2^n s_0, \tag{2.55}$$

$$\sum_{m=1}^{\infty}(-1)^n C_{nm}\xi_1^{m+1}\xi_2^n \gamma_{1n} a_m^{(1)} + 2i^n(2n+1)\sqrt{n(n+1)}\left[x_2 j_n'(x_2) - j_n(x_2)\right]\kappa_{nm}^{(1)}b_m^{(1)}$$
$$+\gamma_{2n} a_n^{(2)} + 2n(n+1)\left[x_2 h_n^{(1)\prime}(x_2) - h_n^{(1)}(x_2)\right]b_n^{(2)} + \gamma_{3n}\widehat{a}_n + \gamma_{4n}\widehat{b}_n$$
$$= i(-1)^{n+1}\omega R_{10}\xi_1\xi_2^n \gamma_{1n} s_0, \tag{2.56}$$

$$\sum_{m=1}^{\infty}2(-1)^{n+1}(n-1)C_{nm}\xi_1^{m+1}\xi_2^n a_m^{(1)} + \frac{i^n(2n+1)}{\sqrt{n(n+1)}}\left[(n^2+n-2)j_n(x_2) + x_2^2 j_n''(x_2)\right]\kappa_{nm}^{(1)}b_m^{(1)}$$
$$2(n+2)a_n^{(2)} + \left[(n^2+n-2)h_n^{(1)}(x_2) + x_2^2 h_n^{(1)\prime\prime}(x_2)\right]b_n^{(2)} + \gamma_{5n}\widehat{a}_n + \gamma_{6n}\widehat{b}_n$$
$$= 2i(-1)^n(n-1)\omega R_{10}\xi_1\xi_2^n s_0, \tag{2.57}$$

where $s_n$ is nonzero only if (2.1) contains a mode with number $n$, and $\gamma_{1n} - \gamma_{6n}$ are defined by

$$\gamma_{1n} = x_2^2 - 2n(n-1), \tag{2.58}$$

$$\gamma_{2n} = x_2^2 - 2(n+1)(n+2), \tag{2.59}$$

$$\gamma_{3n} = \frac{x_l^2}{\eta}\left[\left(i\omega\zeta_p - \lambda_p - \frac{2i\omega\eta_p}{3}\right)j_n(x_l) + 2(\mu_p - i\omega\eta_p)j_n''(x_l)\right], \tag{2.60}$$

$$\gamma_{4n} = 2n(n+1)\frac{\mu_p - i\omega\eta_p}{\eta}\left[j_n(x_t) - x_t j_n'(x_t)\right], \tag{2.61}$$

$$\gamma_{5n} = \frac{2(i\omega\eta_p - \mu_p)}{\eta}\left[j_n(x_l) - x_l j_n'(x_l)\right], \tag{2.62}$$

$$\gamma_{6n} = \frac{i\omega\eta_p - \mu_p}{\eta}\left[(n^2+n-2)j_n(x_t) + x_t^2 j_n''(x_t)\right]. \tag{2.63}$$

Equations (2.54) and (2.55) allow one to express $\hat{a}_n$ and $\hat{b}_n$ in terms of $a_n^{(j)}$ and $b_n^{(j)}$,

$$\hat{a}_n = \alpha_n \left\{ \sum_{m=1}^{\infty} (-1)^n nC_{nm}\xi_1^{m+1}\xi_2^n \gamma_{7n} a_m^{(1)} + \frac{i^n n(n+1)(2n+1)}{\sqrt{n(n+1)}} \gamma_{8n}\kappa_{nm}^{(1)} b_m^{(1)} \right.$$

$$\left. + \gamma_{9n} a_n^{(2)} + \gamma_{10n} b_n^{(2)} + i(-1)^n n\omega R_{10}\xi_1\xi_2^n \gamma_{7n} s_0 \right\}, \tag{2.64}$$

$$\hat{b}_n = \beta_n \left\{ \sum_{m=1}^{\infty} (-1)^n C_{nm}\xi_1^{m+1}\xi_2^n [1+in\omega j_n(x_l)\alpha_n\gamma_{7n}] a_m^{(1)} \right.$$

$$+ \frac{i^n(2n+1)}{\sqrt{n(n+1)}} \left[ in(n+1)\omega j_n(x_l)\alpha_n\gamma_{8n} - j_n(x_2) - x_2 j_n'(x_2) \right] \kappa_{nm}^{(1)} b_m^{(1)}$$

$$+ [1+i\omega j_n(x_l)\alpha_n\gamma_{9n}] a_n^{(2)} + \left[ i\omega j_n(x_l)\alpha_n\gamma_{10n} - h_n^{(1)}(x_2) - x_2 h_n^{(1)\prime}(x_2) \right] b_n^{(2)}$$

$$\left. + i(-1)^n \omega R_{10}\xi_1\xi_2^n [1+in\omega j_n(x_l)\alpha_n\gamma_{7n}] s_0 \right\}, \tag{2.65}$$

where

$$\alpha_n = \frac{1}{i\omega\left\{ x_l x_t j_n'(x_l) j_n'(x_t) + j_n(x_t)\left[ x_l j_n'(x_l) - n(n+1) j_n(x_l) \right] \right\}}, \tag{2.66}$$

$$\beta_n = \frac{1}{i\omega\left[ j_n(x_t) + x_t j_n'(x_t) \right]}, \tag{2.67}$$

$$\gamma_{7n} = nj_n(x_t) - x_t j_n'(x_t), \tag{2.68}$$

$$\gamma_{8n} = x_t j_n'(x_t) j_n(x_2) - x_2 j_n'(x_2) j_n(x_t), \tag{2.69}$$

$$\gamma_{9n} = (n+1)\left[ (n+1) j_n(x_t) + x_t j_n'(x_t) \right], \tag{2.70}$$

$$\gamma_{10n} = n(n+1)\left[ x_t j_n'(x_t) h_n^{(1)}(x_2) - x_2 h_n^{(1)\prime}(x_2) j_n(x_t) \right]. \tag{2.71}$$

Substituting (2.64) and (2.65) into (2.56) and (2.57) and combining the resulting equations with (2.52) and (2.53), one obtains the following system of equations:

$$\sum_{m=1}^{\infty} A_{knm}^{(1)} a_m^{(1)} + B_{knm}^{(1)} b_m^{(1)} + A_{knm}^{(2)} a_m^{(2)} + B_{knm}^{(2)} b_m^{(2)} = D_{kn}, \quad k=1,2,3,4, \tag{2.72}$$

where

$$A_{1nm}^{(1)} = (n+1)\delta_{nm} - n\xi_1^{n+m+1}\xi_2 c_1, \tag{2.73}$$

$$B_{1nm}^{(1)} = \delta_{nm} n(n+1) h_n^{(1)}(x_1), \tag{2.74}$$

$$A^{(2)}_{1nm} = (-1)^{m+1} n C_{nm} \xi_1^n \xi_2^{m+1}, \tag{2.75}$$

$$B^{(2)}_{1nm} = i^n (2n+1)\sqrt{n(n+1)} j_n(x_1) \kappa^{(2)}_{nm}, \tag{2.76}$$

$$D_{1n} = i\omega R_{10}\left(n\xi_1^{n+1}\xi_2 c_1 s_0 + s_n\right), \tag{2.77}$$

$$A^{(1)}_{2nm} = 2\left[\delta_{nm}(n+2) - (n-1)\xi_1^{n+m+1}\xi_2 c_1\right], \tag{2.78}$$

$$B^{(1)}_{2nm} = \delta_{nm}\left[(n^2+n-2)h_n^{(1)}(x_1) + x_1^2 h_n^{(1)\prime\prime}(x_1)\right], \tag{2.79}$$

$$A^{(2)}_{2nm} = 2(-1)^{m+1}(n-1)C_{nm}\xi_1^n \xi_2^{m+1}, \tag{2.80}$$

$$B^{(2)}_{2nm} = \frac{i^n(2n+1)}{\sqrt{n(n+1)}}\kappa^{(2)}_{nm}\left[(n^2+n-2)j_n(x_1) + x_1^2 j_n^{\prime\prime}(x_1)\right], \tag{2.81}$$

$$D_{2n} = 2i(n-1)\xi_1^{n+1}\xi_2 \omega R_{10} c_1 s_0, \tag{2.82}$$

$$A^{(1)}_{3nm} = (-1)^n C_{nm}\xi_1^{m+1}\xi_2^n \left\{\gamma_{1n} + \beta_n\gamma_{4n} + n\alpha_n\gamma_{7n}\left[\gamma_{3n} + i\omega j_n(x_l)\beta_n\gamma_{4n}\right]\right\}, \tag{2.83}$$

$$B^{(1)}_{3nm} = \frac{i^n(2n+1)}{\sqrt{n(n+1)}}\kappa^{(1)}_{nm}\left\{2n(n+1)\left[x_2 j_n^\prime(x_2) - j_n(x_2)\right]\right.$$
$$\left.+n(n+1)\alpha_n\gamma_{8n}\left[\gamma_{3n} + i\omega j_n(x_l)\beta_n\gamma_{4n}\right] - \beta_n\gamma_{4n}\left[j_n(x_2) + x_2 j_n^\prime(x_2)\right]\right\}, \tag{2.84}$$

$$A^{(2)}_{3nm} = \delta_{nm}\{\beta_n\gamma_{4n} + \gamma_{2n} + \alpha_n\gamma_{9n}[\gamma_{3n} + i\omega j_n(x_l)\beta_n\gamma_{4n}]\}, \tag{2.85}$$

$$B^{(2)}_{3nm} = \delta_{nm}\left\{\alpha_n\gamma_{3n}\gamma_{10n} + 2n(n+1)\left[x_2 h_n^{(1)\prime}(x_2) - h_n^{(1)}(x_2)\right]\right.$$
$$\left.+\beta_n\gamma_{4n}\left[i\omega j_n(x_l)\alpha_n\gamma_{10n} - h_n^{(1)}(x_2) - x_2 h_n^{(1)\prime}(x_2)\right]\right\}, \tag{2.86}$$

$$D_{3n} = i(-1)^{n+1}\omega R_{10}\xi_1\xi_2^n s_0\left\{\gamma_{1n} + n\alpha_n\gamma_{3n}\gamma_{7n} + \beta_n\gamma_{4n}\left[1 + in\omega j_n(x_l)\alpha_n\gamma_{7n}\right]\right\}, \tag{2.87}$$

$$A^{(1)}_{4nm} = (-1)^n C_{nm}\xi_1^{m+1}\xi_2^n\left\{\beta_n\gamma_{6n} - 2(n-1) + n\alpha_n\gamma_{7n}\left[\gamma_{5n} + i\omega j_n(x_l)\beta_n\gamma_{6n}\right]\right\}, \tag{2.88}$$

$$B^{(1)}_{4nm} = \frac{i^n(2n+1)}{\sqrt{n(n+1)}}\kappa^{(1)}_{nm}\left\{(n^2+n-2)j_n(x_2) + x_2^2 j_n^{\prime\prime}(x_2)\right.$$
$$\left.+n(n+1)\alpha_n\gamma_{8n}\left[\gamma_{5n} + i\omega j_n(x_l)\beta_n\gamma_{6n}\right] - \beta_n\gamma_{6n}\left[j_n(x_2) + x_2 j_n^\prime(x_2)\right]\right\}, \tag{2.89}$$

$$A^{(2)}_{4nm} = \delta_{nm}\left\{2(n+2) + \alpha_n\gamma_{5n}\gamma_{9n} + \beta_n\gamma_{6n}\left[1 + i\omega j_n(x_l)\alpha_n\gamma_{9n}\right]\right\}, \tag{2.90}$$

$$B^{(2)}_{4nm} = \delta_{nm}\left\{(n^2+n-2)h_n^{(1)}(x_2) + x_2^2 h_n^{(1)\prime\prime}(x_2) + \alpha_n\gamma_{10n}\left[\gamma_{5n} + i\omega j_n(x_l)\beta_n\gamma_{6n}\right]\right.$$
$$\left.-\beta_n\gamma_{6n}\left[h_n^{(1)}(x_2) + x_2 h_n^{(1)\prime}(x_2)\right]\right\}, \tag{2.91}$$

$$D_{4n} = i(-1)^n \omega R_{10} \xi_1 \xi_2^n s_0 \{2(n-1) - \beta_n \gamma_{6n} - n\alpha_n \gamma_{7n}[\gamma_{5n} + i\omega j_n(x_l)\beta_n \gamma_{6n}]\}. \tag{2.92}$$

Equation (2.72) is a system with an infinite number of equations, which contain an infinite number of terms. However, since the scattering coefficients decrease with increasing scattering order, (2.72) can be truncated at $n = m = N$. Doing so in (2.72), we obtain 4$N$ equations in 4$N$ unknowns $a_n^{(1)}$, $b_n^{(1)}$, $a_n^{(2)}$ and $b_n^{(2)}$ with $1 \leq n \leq N$. This system can be solved numerically. Changing $N$, the scattering coefficients can be calculated with any desired accuracy.

## 2.5. *Acoustic streaming*

To model acoustic streaming in the case of a bubble and a particle, we can use equations derived by Doinikov *et al.* (2022) for the case of two bubbles, substituting the linear scattering coefficients calculated in the present study. However, we need to re-calculate constants that appear in those equations. They are denoted by $C_{1l0}^{(j)}$, $C_{2l0}^{(j)}$, $C_{3l0}^{(j)}$, $C_{4l0}^{(j)}$ with $j = 1,2$. These constants are calculated by boundary conditions at the surfaces of two bubbles, which require that the normal velocity and the tangential stress of the Lagrangian streaming vanish at the equilibrium bubble surfaces. In the case of a bubble and a particle, we keep the same conditions for the bubble but we change one condition for the particle. The condition of zero tangential stress should be replaced by the condition of zero tangential velocity of the Lagrangian streaming at the equilibrium surface of the particle. This condition is given by

$$v_{L\theta}(r_2, \theta_2) = 0 \text{ at } r_2 = R_{20}, \tag{2.93}$$

where $v_{L\theta}(r_2, \theta_2) = v_{E\theta}(r_2, \theta_2) + v_{S\theta}(r_2, \theta_2)$ is the tangential component of the Lagrangian streaming velocity, $v_{E\theta}(r_2, \theta_2)$ is the tangential component of the Eulerian streaming velocity and $v_{S\theta}(r_2, \theta_2)$ is the tangential component of the Stokes drift velocity (Doinikov *et al.* 2022).

The velocities $v_{E\theta}(r_2, \theta_2)$ and $v_{S\theta}(r_2, \theta_2)$ are given by (C16) and (E16) of (Doinikov *et al.* 2022). Substitution of these equations into (2.93) yields

$$\Psi_l^{(2)}(R_{20}) + R_{20}\Psi_l^{(2)\prime}(R_{20}) = R_{20}V_{S\theta l}^{(2)}(R_{20}), \tag{2.94}$$

where $\Psi_l^{(2)}$, $\Psi_l^{(2)\prime}$ and $V_{S\theta l}^{(2)}$ are defined by (C8), (C17) and (E20) of (Doinikov *et al.* 2022). $\Psi_l^{(2)}$ and $\Psi_l^{(2)\prime}$ contain the desired constants $C_{1l0}^{(j)} - C_{4l0}^{(j)}$, while $V_{S\theta l}^{(2)}(R_{20})$ is a known function that is expressed in terms of $a_n^{(j)}$ and $b_n^{(j)}$.

Substitution of $\Psi_l^{(2)}$ and $\Psi_l^{(2)\prime}$ into (2.94) yields

$$(l-2)C_{1l0}^{(2)} + \frac{lC_{2l0}^{(2)}}{R_{20}^2} - (l+1)R_{20}^{2l-1}C_{3l0}^{(2)} - (l+3)R_{20}^{2l+1}C_{4l0}^{(2)} = X_{2l}^{(p)}, \qquad (2.95)$$

where

$$X_{2l}^{(p)} = -R_{20}^l V_{S\theta l}^{(2)}(R_{20}). \qquad (2.96)$$

Equation (2.95) replaces (C20) of (Doinikov *et al.* 2022) at $j = 2$.

We also need to replace (C28) of (Doinikov *et al.* 2022) at $j = 1$, using the condition of zero tangential velocity at the particle surface instead of the condition of zero tangential stress. This condition requires that

$$v_{L\theta}(r_1, \theta_1) = v_{E\theta}(r_1, \theta_1) + v_{S\theta}(r_1, \theta_1) = 0 \quad \text{at} \quad \theta_1 = 0, \; r_1 = d - R_{20}. \qquad (2.97)$$

Substituting $v_{E\theta}(r_1, \theta_1)$ and $v_{S\theta}(r_1, \theta_1)$, given by (C16) and (E16) of (Doinikov *et al.* 2022), into (2.97), one obtains

$$(l-2)C_{1l0}^{(1)} + \frac{lC_{2l0}^{(1)}}{(d-R_{20})^2} - (l+1)(d-R_{20})^{2l-1}C_{3l0}^{(1)} - (l+3)(d-R_{20})^{2l+1}C_{4l0}^{(1)} = X_{4l}^{(p)}, \qquad (2.98)$$

where

$$X_{4l}^{(p)} = -(d-R_{20})^l V_{S\theta l}^{(1)}(d-R_{20})$$

$$-\frac{1}{2(2l-1)(2l+1)} \int_{R_{10}}^{d-R_{20}} \left[ (l-2)s^{l+2} + (l+1)(d-R_{20})^{2l-1}s^{3-l} \right] E_l^{(1)}(s)\,ds$$

$$+\frac{1}{2(2l+1)(2l+3)} \int_{R_{10}}^{d-R_{20}} \left[ \frac{ls^{l+4}}{(d-R_{20})^2} + \frac{(l+3)(d-R_{20})^{2l+1}}{s^{l-1}} \right] E_l^{(1)}(s)\,ds \qquad (2.99)$$

and $E_l^{(1)}(s)$ is given by (C5) of (Doinikov *et al.* 2022). Equation (2.98) replaces (C28) of (Doinikov *et al.* 2022) at $j = 1$.

Combining (2.95) and (2.98) with the following equations of (Doinikov *et al.* 2022): (C19) at $j = 1, 2$, (C20) at $j = 1$, (C23), (C25) and (C28) at $j = 2$, one obtains the following expressions for the desired constants:

$$C_{1l0}^{(1)} = \frac{l(l+2)X_{1l}^{(1)}}{2l+1} - \frac{X_{2l}^{(1)}}{2l+1} - R_{10}^{2l-1}C_{3l0}^{(1)}, \qquad (2.100)$$

$$C_{2l0}^{(1)} = \frac{(1-l^2)R_{10}^2 X_{1l}^{(1)}}{2l+1} + \frac{R_{10}^2 X_{2l}^{(1)}}{2l+1} - R_{10}^{2l+3}C_{4l0}^{(1)}, \qquad (2.101)$$

$$C_{3l0}^{(1)} = \frac{1}{(d-R_{20})^{2l-1} - R_{10}^{2l-1}} \left\{ \frac{X_{1l}^{(1)}[(l^2-1)R_{10}^2 - l(l+2)(d-R_{20})^2]}{(2l+1)(d-R_{20})^2} \right.$$

$$\left. + \frac{X_{2l}^{(1)}[(d-R_{20})^2 - R_{10}^2]}{(2l+1)(d-R_{20})^2} + X_{3l}^{(1)} + \frac{R_{10}^{2l+3} - (d-R_{20})^{2l+3}}{(d-R_{20})^2} C_{4l0}^{(1)} \right\}, \quad (2.102)$$

$$C_{4l0}^{(1)} = \frac{(d-R_{20})^2}{2(d-R_{20})^{4l+2} - 2R_{10}^{4l+2} + (2l+1)R_{10}^{2l-1}(d-R_{20})^{2l-1}\left[R_{10}^4 - (d-R_{20})^4\right]}$$

$$\times \left\{ \frac{X_{1l}^{(1)}[l(l+2)(2l-1)(d-R_{20})^{2l+1} - (l^2-1)(2l+1)R_{10}^2(d-R_{20})^{2l-1} + 2(l^2-1)R_{10}^{2l+1}]}{(2l+1)(d-R_{20})^2} \right.$$

$$+ \frac{X_{2l}^{(1)}\left[(2l+1)R_{10}^2(d-R_{20})^{2l-1} - (2l-1)(d-R_{20})^{2l+1} - 2R_{10}^{2l+1}\right]}{(2l+1)(d-R_{20})^2}$$

$$\left. -X_{3l}^{(1)}\left[(l-2)R_{10}^{2l-1} + (l+1)(d-R_{20})^{2l-1}\right] - X_{4l}^{(p)}\left[(d-R_{20})^{2l-1} - R_{10}^{2l-1}\right] \right\}, \quad (2.103)$$

$$C_{1l0}^{(2)} = \frac{1}{2}\left[lX_{1l}^{(2)} - X_{2l}^{(p)} - (2l+1)R_{20}^{2l-1}C_{3l0}^{(2)} - (2l+3)R_{20}^{2l+1}C_{4l0}^{(2)}\right], \quad (2.104)$$

$$C_{2l0}^{(2)} = \frac{1}{2}\left[(2-l)R_{20}^2 X_{1l}^{(2)} + R_{20}^2 X_{2l}^{(p)} + (2l-1)R_{20}^{2l+1}C_{3l0}^{(2)} + (2l+1)R_{20}^{2l+3}C_{4l0}^{(2)}\right], \quad (2.105)$$

$$C_{3l0}^{(2)} = \left[2(d-R_{10})^{2l+1} - (2l+1)R_{20}^{2l-1}(d-R_{10})^2 + (2l-1)R_{20}^{2l+1}\right]^{-1}$$

$$\times \left\{ X_{1l}^{(2)}\left[(l-2)R_{20}^2 - l(d-R_{10})^2\right] + X_{2l}^{(p)}\left[(d-R_{10})^2 - R_{20}^2\right] + 2(d-R_{10})^2 X_{3l}^{(2)} \right.$$

$$\left. + \left[(2l+3)R_{20}^{2l+1}(d-R_{10})^2 - 2(d-R_{10})^{2l+3} - (2l+1)R_{20}^{2l+3}\right]C_{4l0}^{(2)} \right\}, \quad (2.106)$$

$$C_{4l0}^{(2)} = \left\{(l^2 + 2l - q_l)\left[2(d-R_{10})^{2l+3} + (2l+1)R_{20}^{2l+3}\right] + (2l+3)(q_l - l^2 + 1)R_{20}^{2l+1}(d-R_{10})^2\right\}^{-1}$$

$$\times \left\{ X_{1l}^{(2)}\left\{\left[l(l^2-4) - (l-2)q_l\right]R_{20}^2 + l\left(q_l - l^2 + 1\right)(d-R_{10})^2\right\} \right.$$

$$+ X_{2l}^{(p)}\left[(l^2 - 1 - q_l)(d-R_{10})^2 + (q_l - l^2 - 2l)R_{20}^2\right]$$

$$\left. -2q_l(d-R_{10})^2 X_{3l}^{(2)} + 2(d-R_{10})^2 X_{4l}^{(2)} \right\}, \quad (2.107)$$

where

$$q_l = \frac{2(l^2-1)(d-R_{10})^{2l+1} - (l^2-1)(2l+1)R_{20}^{2l-1}(d-R_{10})^2 + l(l+2)(2l-1)R_{20}^{2l+1}}{2(d-R_{10})^{2l+1} - (2l+1)R_{20}^{2l-1}(d-R_{10})^2 + (2l-1)R_{20}^{2l+1}} \quad (2.108)$$

and $X_{1l}^{(j)}$, $X_{2l}^{(1)}$, $X_{3l}^{(j)}$ and $X_{4l}^{(2)}$ are defined by (C21), (C22), (C24), (C26) and (C29) of (Doinikov et al. 2022).

Equations (2.100) – (2.107) replace (C30) – (C33) of (Doinikov et al. 2022). All the other solutions obtained by Doinikov et al. (2022) for acoustic streaming remain the same.

### 2.6. Shear stress around the particle

The time-averaged shear stress in the liquid around the particle is given by

$$\sigma_{Lr\theta}(r_2,\theta_2) = \eta\left[\frac{1}{r_2}\frac{\partial v_{Lr}(r_2,\theta_2)}{\partial \theta_2} + \frac{\partial v_{L\theta}(r_2,\theta_2)}{\partial r_2} - \frac{v_{L\theta}(r_2,\theta_2)}{r_2}\right], \quad (2.109)$$

where $v_{Lr}(r_2,\theta_2) = v_{Er}(r_2,\theta_2) + v_{Sr}(r_2,\theta_2)$ and $v_{L\theta}(r_2,\theta_2) = v_{E\theta}(r_2,\theta_2) + v_{S\theta}(r_2,\theta_2)$ are calculated by (C15), (C16), (E15) and (E16) of (Doinikov et al. 2022). Substitution of these equations into (2.109) yields

$$\sigma_{Lr\theta}(r_2,\theta_2) = \eta\sum_{l=1}^{\infty} P_l^1(\mu_2)\left[\frac{(2-l^2-l)\Psi_l^{(2)}(r_2)}{r_2^2} - \Psi_l^{(2)//}(r_2) + \frac{V_{Srl}^{(2)}(r_2) - V_{S\theta l}^{(2)}(r_2)}{r_2} + V_{S\theta l}^{(2)/}(r_2)\right], \quad (2.110)$$

where $\Psi_l^{(2)}(r_2)$, $\Psi_l^{(2)//}(r_2)$, $V_{Srl}^{(2)}(r_2)$, $V_{S\theta l}^{(2)}(r_2)$ and $V_{S\theta l}^{(2)/}(r_2)$ are calculated by (C8), (C18), (E19), (E20) and (E21) of (Doinikov et al. 2022). Substitution of these equations into (2.110) yields

$$\sigma_{Lr\theta}(r_2,\theta_2) = \eta\sum_{l=1}^{\infty} P_l^1(\mu_2)\left\{2(1-l^2)\left[\frac{C_{1l}^{(2)}(r_2)}{r_2^{l+1}} + r_2^{l-2}C_{3l}^{(2)}(r_2)\right] - l(l+2)\left[\frac{2C_{2l}^{(2)}(r_2)}{r_2^{l+3}} + r_2^l C_{4l}^{(2)}(r_2)\right]\right.$$

$$\left. + \frac{V_{Srl}^{(2)}(r_2) - V_{S\theta l}^{(2)}(r_2)}{r_2} + V_{S\theta l}^{(2)/}(r_2)\right\}, \quad (2.111)$$

where the functions $C_{1l}^{(2)}(r_2) - C_{4l}^{(2)}(r_2)$ are calculated by (C9) – (C12) of (Doinikov et al. 2022).

On the particle surface, the shear stress is given by

$$\sigma_{Lr\theta}(R_{20},\theta_2) = \eta\sum_{l=1}^{\infty} P_l^1(\mu_2)\left[2(1-l^2)\left(\frac{C_{1l0}^{(2)}}{R_{20}^{l+1}} + R_{20}^{l-2}C_{3l0}^{(2)}\right) - l(l+2)\left(\frac{2C_{2l0}^{(2)}}{R_{20}^{l+3}} + R_{20}^l C_{4l0}^{(2)}\right)\right.$$

$$\left. + \frac{V_{Srl}^{(2)}(R_{20}) - V_{S\theta l}^{(2)}(R_{20})}{R_{20}} + V_{S\theta l}^{(2)/}(R_{20})\right], \quad (2.112)$$

where the constants $C_{1l0}^{(2)} - C_{4l0}^{(2)}$ are calculated by (2.104) – (2.107).

## 3. Numerical examples

Figure 2 illustrates acoustic streaming produced by different oscillation modes, which are experienced by the bubble, and the time-averaged shear stress produced by these modes on the particle surface. The liquid surrounding the bubble and the particle is assumed to be water with $\rho = 1000$ kg/m$^3$ and $\eta = 0.001$ Pa s. The particle material is steel with the following parameters: $\rho_p = 7800$ kg/m$^3$, $E = 206$ GPa, $\sigma = 0.28$, $\eta_p = 0$ and $\zeta_p = 0$. The bubble and the particle radii are $R_{10} = R_{20} = 5$ µm and the distance between the centres of the bubble and the particle is $d = 20$ µm, which means that the distance between the bubble surface and the particle surface is $2R_{10}$. Figures 2(a,b), 2(c,d) and 2(e,f) show the acoustic streaming and the stress in the cases that the bubble undergoes the radial pulsation (mode 0), the translational oscillation (mode 1) and both of the above modes (modes 0+1). The insert in figure 2(b) illustrates the distribution and the direction of the stress on the particle surface. Note that the red line in the insert tries to reflect the fact that the stress curve in the range $\pi \leq \theta_2 \leq 2\pi$ is the inverted image of the stress curve in the range $0 \leq \theta_2 \leq \pi$, i.e., that the stress for $\theta_2 \geq \pi$ has the same magnitude as the stress for $\theta_2 \leq \pi$ but the opposite sign. The oscillation frequency is $f = 500$ kHz and the mode amplitudes are $s_0 = s_1 = 1$ µm. This choice of the mode amplitudes, which we will also follow in other examples, is convenient for the following reason: if the modes have amplitudes of $a$ µm and $b$ µm, it is enough to multiply our results by the product $ab$.

In the case shown in figure 2(a), where the bubble undergoes the radial pulsation alone, the spatial structure of the streaming is constituted of four recirculation loops surrounding the bubble and of four small vortices around the particle that are located on the side facing the bubble. It is worth reminding that a single (i.e. located in an unbounded liquid) radially pulsating bubble does not generate acoustic streaming (Doinikov *et al.* 2019). Therefore, the presence of a (here rigid) particle is sufficient to significantly modify the structure of the flow motion. The case shown in figure 2(c), where the bubble undergoes the translational oscillation alone, demonstrates a typical quadrupole-like streaming pattern investigated by Longuet-Higgins (1998). The same recirculation loops structure is observed around the particle. In view of the symmetry of the vortices around the bubble from the left-to-right side, it is clear that the particle does not influence strongly the flow motion in the case of a bubble experiencing a translation oscillation alone. Figure 2(e) shows the streaming pattern that results from combining the radial and translational oscillations. The streaming around the bubble strongly differs from the dipole-like structure predicted for a single bubble undergoing modes 0 and 1 (Longuet-Higgins 1998; Doinikov *et al.* 2019). The streaming around the bubble shown in figure 2(e), with the four asymmetric vortices along the *z* direction, highlights a strong influence of the neighboring particle.

Figures 2(b), 2(d) and 2(f) show that the maximal stress magnitude is reached on the particle side that faces the bubble, near the point with $\theta_2 = \pi$. Note that in view of the radial symmetry, the stress is zero at the point $\theta_2 = \pi$. However, if we take a circle around the above point, we will see that the stress along the perimeter of this circle is directed away from the center. This means that the particle surface inside the circle undergoes stretching forces directed away from the center, which tear up the particle surface (see the insert in figure 2(b)). Considering the associated microstreaming pattern, this stretching is related to the existence of two small counter-rotating vortices at $\theta_2 = \pi$. As a result, the surface rupture (hole) can appear at $\theta_2 = \pi$. Therefore, the point $\theta_2 = \pi$ is a possible point of rupture.

Figures 2(g) and 2(h) show the streaming and the stress in the case that the bubble undergoes mode 2 assuming that $f = 250$ kHz and $s_2 = 1$ µm. This choice of the frequency implies that mode 2 is excited parametrically by the radial 500-kHz mode. The streaming pattern around the bubble is constituted of eight recirculation loops and resembles that for a single bubble undergoing mode 2 (Inserra *et al.* 2020). It is interesting to note that small counter-rotating vortices near the particle surface at $\theta_2 = \pi$ are not visible. The particle interface is however teared at this location but, as one can see, the shear stress gradient is small. One can therefore hypothesize that the rupture probability at $\theta_2 = \pi$ is small. The maximum shear stress now occurs near the location $\theta_2 = 0$, which becomes the most probable location for rupture. The comparison of the stress curves in figures 2(b), 2(d), 2(f) and 2(h) shows that the stress magnitude in the case of a single translational or nonspherical oscillation is the smallest. The highest stress magnitudes are observed when the bubble oscillation contains mode 0. This fact is well known for a single bubble (Longuet-Higgins 1998) and here it is also valid for the shear stress on the particle surface.

One comment on the numerical calculation of the stress should be made. As one can see, the upper limit of the sum in (2.112) is infinite. In the numerical calculation, we set a finite upper limit and increase it as long as the stress curve does not change any more. In figures 2(b), 2(d), 2(f) and 2(h), the stress curves do not change any more when the upper limit of the sum in (2.112) reaches 11, 10, 9 and 7, respectively. The process of the convergence of the stress curve to the final result can be characterized by the following quantity:

$$s_l = \sum_k \left[ \sigma_{Lr\theta}^{(l)}(R_{20}, \theta_{2k}) - \sigma_{Lr\theta}^{(\infty)}(R_{20}, \theta_{2k}) \right]^2, \qquad (3.1)$$

where $0 \leq \theta_{2k} \leq \pi$, $\sigma_{Lr\theta}^{(l)}(R_{20}, \theta_{2k})$ is calculated by (2.112) setting the upper limit of the sum equal to $l$ and $\sigma_{Lr\theta}^{(\infty)}(R_{20}, \theta_{2k})$ is the value of the stress when the stress curve does not change any more. As an example, figure 3 shows $s_l$ for the stress curve presented in figure 2(b).

Figure 4 illustrates cases where the particle material is different from steel. Figure 4(a) shows the acoustic streaming assuming that the particle material is copolymer with $\rho_p = 900$ kg/m³, $E = 1.1$ GPa, $\sigma = 0.42$, $\eta_p = 0$ and $\zeta_P = 0$. Figure 4(b) shows the streaming assuming that the particle material is polystyrene with $\rho_p = 1050$ kg/m³, $E = 3.4$ GPa, $\sigma = 0.34$, $\eta_p = 0$ and $\zeta_P = 0$. These materials are of interest because they are widely used in microfluidics (Leibacher *et al*. 2015; Nilsson *et al*. 2009). Figure 4(c) illustrates the case where the material parameters of the particle correspond to erythrocytes with $\rho_p = 1125$ kg/m³, $E = 26$ kPa, $\sigma = 0.49$, $\eta_p = 0.006$ Pa s and $\zeta_P = 0$ (Dulińska *et al*. 2006). In all these cases, the bubble is assumed to undergo modes 0 and 1. For this excitation, the shear stress is shown in figure 2 to be the highest. The other parameters are as in figure 2. The stress curves for the above cases are shown in figure 4(d). They were obtained setting the upper limit of the sum in (2.112) equal to 9, 9 and 8, respectively. For comparison, the stress curve for the steel particle, shown in figure 2(f), is also included in figure 4(d).

For copolymer and polystyrene (materials with high rigidity), the overall streaming pattern is similar to that for the steel particle, shown in figure 2(e). The counter-rotating vortices are observed near the particle at $\theta_2 = \pi$ and result in the rupture point on the particle surface at this location. The second possible rupture location is around the equator at $\theta_2 = \pi/2$, where another recirculation loop is clearly visible. A significant change occurs when a softer particle (erythrocyte) is considered. The direction of the streaming flow reverses around the location $\theta_2 = 0$ in comparison to the other stiff materials, and the recirculation vortices around the point facing the bubble almost disappear. In addition, while the distribution of the shear stress along the interface of the erythrocyte-like particle is similar to that for the stiffer particles, the stress magnitude for the erythrocyte-like particle is much smaller. Also, while the two poles of the particle are the places where rupture can occur in the case of a steel particle, softer materials (copolymer, polystyrene or erythrocyte) can only be ruptured near the equator or at the point facing the bubble. These findings differ from available results on bubble-induced mechanical effects (for instance, in the context of bubble-induced cell poration), where the cell is considered as an infinite elastic wall in front of which a single, infinitely small bubble is oscillating. For such a configuration, a single location of rupture is found at the point of the cell (wall) interface facing the bubble. When a finite-sized particle rather than an infinite wall is considered, the results of figures 2,4,5 display new findings, with the possibility of new points of rupture on the particle interface. When the bubble is experiencing the radial mode 0, the combination of radial and translational oscillations 0-1 or the translational interaction 1-1 (figure 2(a-c)) near a steel particle, the point of the particle interface facing the bubble is always the most probable location for rupture (with the highest stress and stress gradient magnitudes). For the same bubble driving, this behavior does not change for stiff particles (copolymer and polystyrene, figure 4(d)), but a change occurs for the erythrocyte-like particle. The shear stress gradient near the equator ($\theta = \pi/2$) becomes similar in amplitude to that at the point facing the bubble ($\theta = \pi$), so the equator becomes another location where rupture can be predominant.

Figure 5 illustrates the effect of the particle viscoelasticity. The parameters used, except the particle shear viscosity $\eta_p$, are as in figure 4(c) for the erythrocyte-like particle. Figures 5(a) and 5(b) show acoustic streaming for $\eta_p = 0.006$ Pa s and $\eta_p = 0.06$ Pa s, respectively. Figure 5(c) shows the shear stress on the particle surface for increasing values of $\eta_p$: 0.006 Pa s, 0.012 Pa s, 0.03 Pa s and 0.06 Pa s. As one can see, while the microstreaming pattern around the bubble remains almost unchanged whatever the particle shear viscosity (cf. figures 5(a) and 5(b)), additional recirculation vortices appear near the particle for the higher shear viscosity. The distribution of the shear stress on the particle surface follows the same trend when $\eta_p$ increases (figure 5(c)) but the maximal shear stress increases as the viscoelastic properties of the particle increase. Furthermore, the location of possible rupture at the particle equator and at the pole facing the bubble remains unchanged when the cell viscoelasticity is modified. The stress gradient at the point facing the bubble significantly increases, suggesting that high-viscosity particles may be more subjected to rupture under similar acoustic conditions.

Figure 6(b) shows a stress curve that was calculated for a radially oscillating bubble near a big particle at $R_{10} = 5$ µm, $R_{20} = 50$ µm, $d = 65$ µm, $f = 500$ kHz and $s_0 = 1$ µm, the other parameters being as in figure 2. The curve does not change any more when the upper limit of the sum in (2.112) reaches 35. Figure 6(a) shows the system under consideration, keeping the real ratio of all sizes. Since $R_{10} \ll R_{20}$, this case can be considered as an approximation of the behaviour of a bubble near a plane rigid wall. Note that, mathematically, "rigid wall" means that Young's modulus and the density of the wall tend to infinity. Numerical simulations do not allow infinite parameters. Therefore, for definiteness, we take Young's modulus and the density of steel. They are quite high, which allows one to consider that the behavior of a steel wall is close to that of a rigid wall.

The time-averaged shear stress produced by an oscillating bubble on a plane rigid wall is commonly evaluated by using an approximation proposed by Nyborg (1958) for calculating acoustic streaming near a boundary. Based on this approximation, Doinikov & Bouakaz (2014) have derived the following formula:

$$\sigma_\tau = \frac{\rho \delta}{8} \frac{\partial}{\partial \tau} \left( v_{\tau 0} v_{\tau 0}^* \right). \tag{3.2}$$

where $\sigma_\tau$ is the time-averaged shear stress produced by a radially oscillating bubble on a plane rigid wall, $\tau$ is the distance from the $z$ axis along the wall and $v_{\tau 0}$ is the first-order velocity component directed along the wall (along $\tau$), which corresponds to the potential liquid flow and is taken at $z = d_w$, where $d_w$ is the distance between the bubble center and the wall; see figure 7(a). In the designations used in the present paper, $v_{\tau 0}$ is represented by

$$v_{\tau 0} = -e^{-i\omega t} \frac{2i\omega R_{10}^2 s_0 \tau}{(d_w^2 + \tau^2)^{3/2}} \left[ 1 + \frac{R_{10}^3 \chi}{d_w (d_w^2 + \tau^2)} \right], \tag{3.3}$$

where

$$\chi = \frac{3(x_1^3 + 3ix_1^2 - 6x_1 - 6i)}{4(18i + 18x_1 - 3ix_1^2 - x_1^3)}. \tag{3.4}$$

Substitution of (3.3) into (3.2) yields

$$\sigma_\tau = \frac{\omega^2 \rho \delta R_{10}^4 |s_0|^2 \tau}{(d_w^2 + \tau^2)^4} \left[ d_w^2 - 2\tau^2 + \frac{2R_{10}^3 \operatorname{Re}\{\chi\}(d_w^2 - 3\tau^2)}{d_w(d_w^2 + \tau^2)} + \frac{R_{10}^6 |\chi|^2 (d_w^2 - 4\tau^2)}{d_w^2(d_w^2 + \tau^2)^2} \right]. \quad (3.5)$$

For the same parameters as in figure 6, (3.5) gives the stress curve shown in figure 7(b). Comparing figures 6(b) and 7(b), one can see that the maximal stress magnitude predicted by our theory is 25.2 Pa, whereas (3.5) gives 1.1 Pa. This comparison suggests that the evaluation of the shear stress based on Nyborg's approximation may considerably underestimate the stress magnitude.

In figures 6 and 7, the distance between the bubble surface and the particle surface (wall) is $2R_{10} = 10$ µm. We have also made calculations for smaller distances between the bubble surface and the particle surface (wall): 5 µm and 1 µm. For these cases, the maximal stress magnitude predicted by our theory is 155.4 Pa and 534.5 Pa, respectively, whereas (3.5) gives 7.4 Pa and 49.7 Pa, respectively. These results confirm that the calculation based on Nyborg's approximation predicts a much lower stress than our model, which is based on a more rigourous calculation of acoustic streaming. It should be noted, however, that the difference between the results given by (3.5) and our model decreases with decreasing distance between the bubble and the particle.

## 4. Conclusion

We have developed an analytical theory that describes acoustic microstreaming produced by the interaction of an oscillating gas bubble with a viscoelastic particle at an arbitrary separation distance between them. The bubble was assumed to undergo axisymmetric oscillation modes. The oscillations of the particle were assumed to be excited by the oscillations of the bubble. In order to demonstrate the capabilities of the developed theory, we have presented numerical examples, which show acoustic streaming produced by different oscillation modes experienced by the bubble and the time-averaged shear stress produced by this streaming on the particle surface. In particular, it has been shown that the stress predicted by our theory is much higher than the stress predicted by Nyborg's approximation (Nyborg 1958), which is commonly used to evaluate the time-averaged shear stress produced by a bubble on a rigid wall.

**Funding.** This research was funded by l'Agence Nationale de la Recherche (ANR) under the project ANR-22-CE92-0062.
**Declaration of interest.** The authors report no conflict of interest.

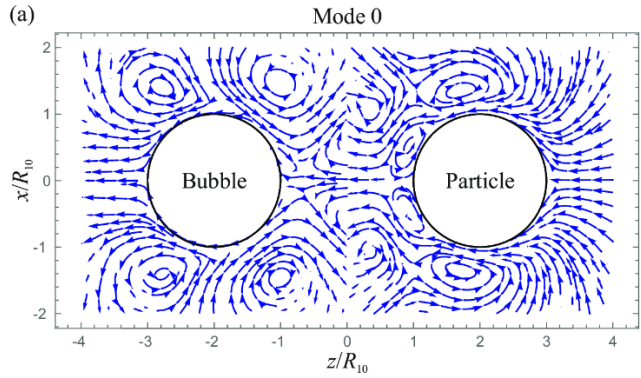
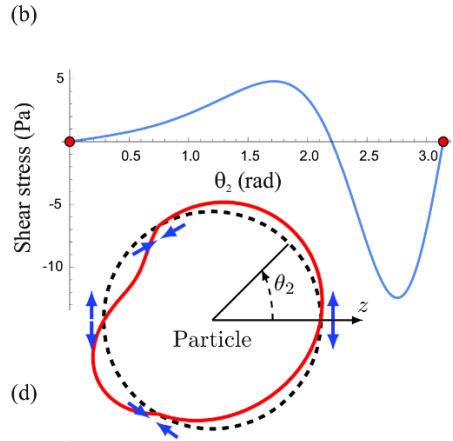
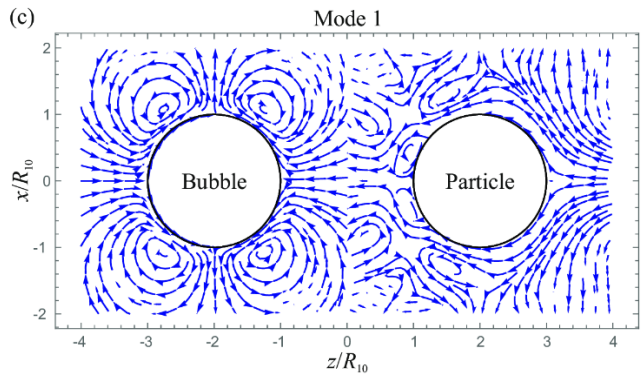
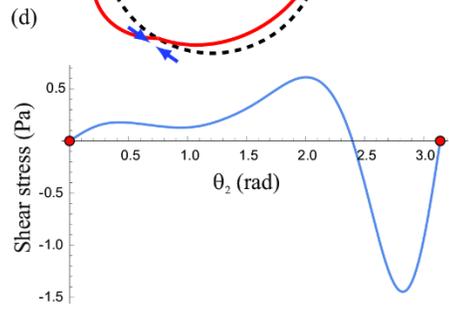
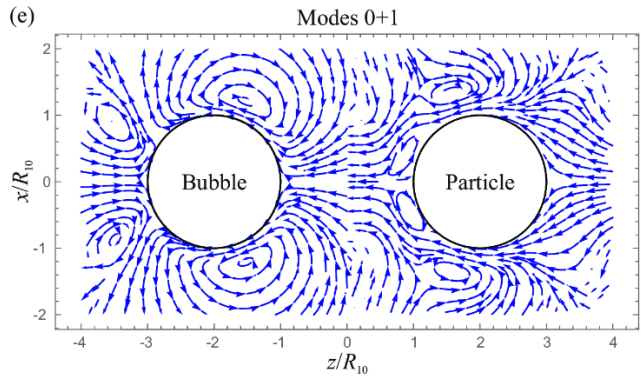
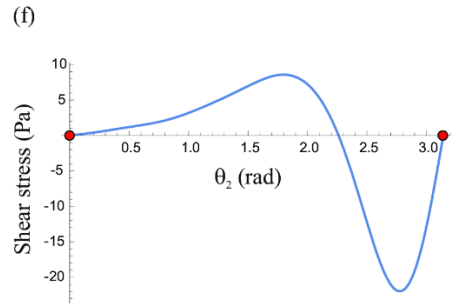
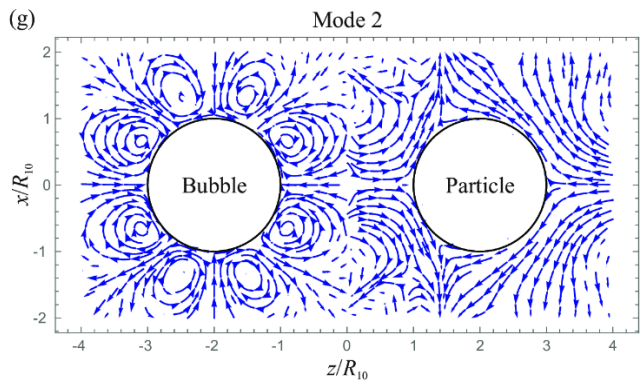
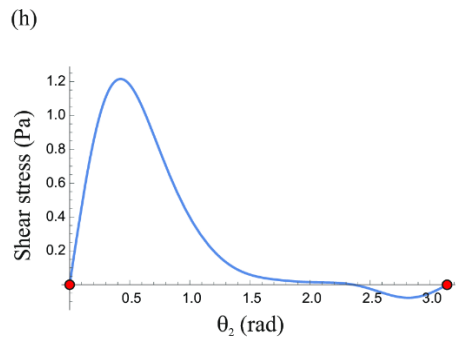

Figure 2. Acoustic streaming (left) and shear stress on the particle surface (right) produced by different oscillation modes experienced by the bubble. The surrounding liquid is water, the particle material is steel, $R_{10} = R_{20} = 5$ µm, $d = 20$ µm, $s_0 = s_1 = s_2 = 1$ µm, the oscillation frequency is (a) – (f) 500 kHz, (g), (h) 250 kHz. Possible places of rupture are shown by small full circles on the stress curves. The insert in (b) displays the evolution of the shear stress along the particle interface, as well as the locations of the stretching and compression points on the particle surface.

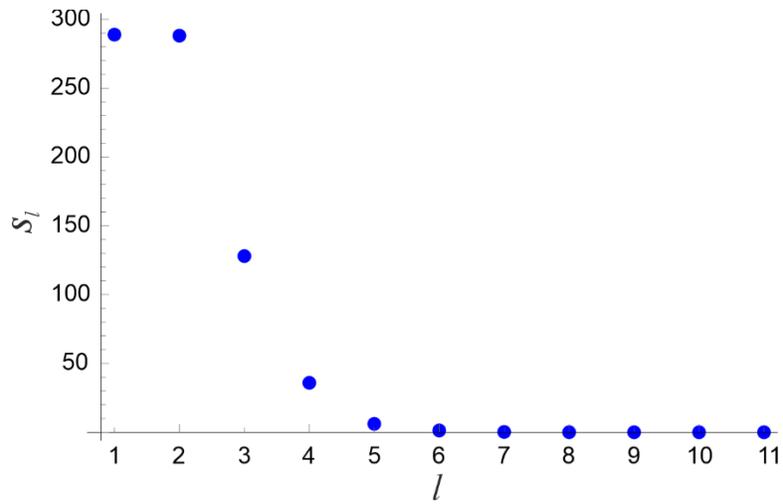

Figure 3. Calculation of the convergence $s_l$ for the stress curve shown in figure 2(b).

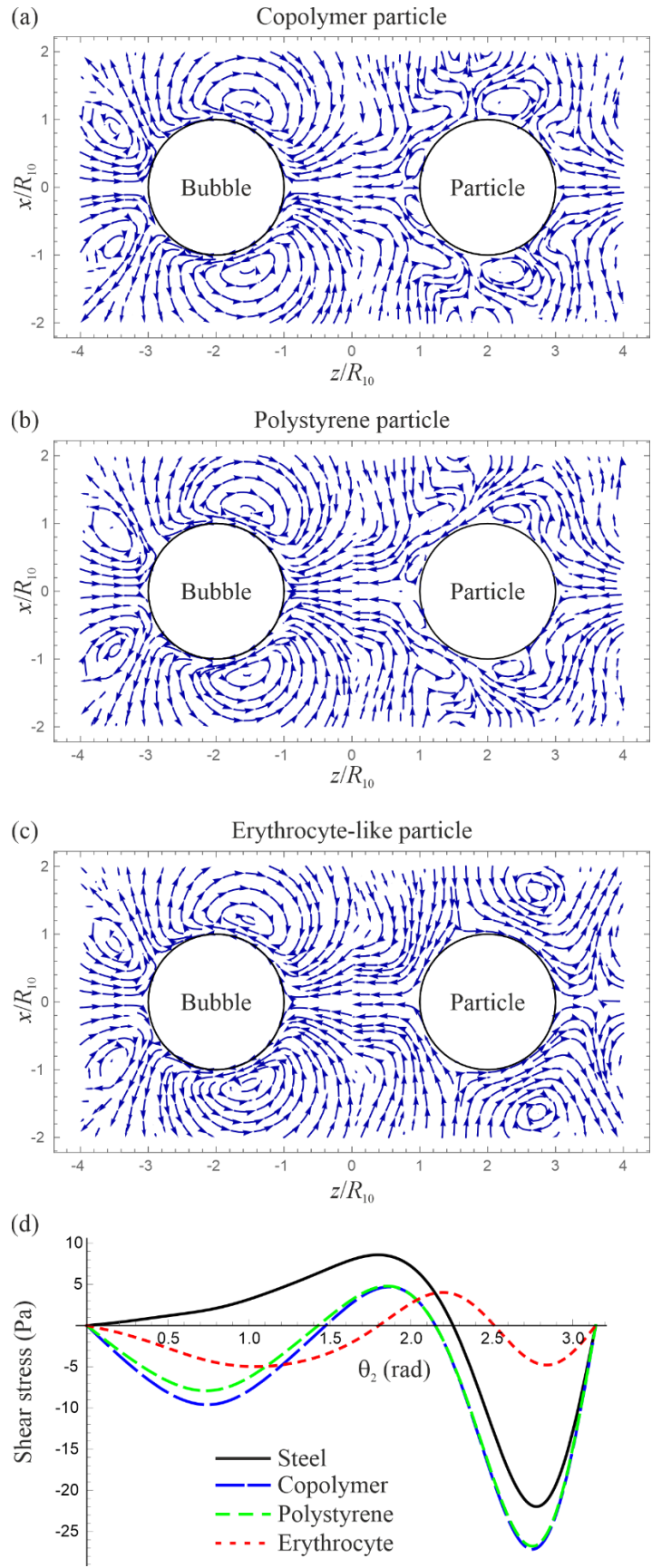

Figure 4. Acoustic streaming in the case that the particle material is (a) copolymer, (b) polystyrene and (c) the material parameters of the particle correspond to erythrocytes. The bubble undergoes modes 0 and 1. The other parameters are as in figure 2. (d) The stress curves for the cases shown in figures 4(a) – 4(c) and for the steel particle shown in figures 2(e) and 2(f).

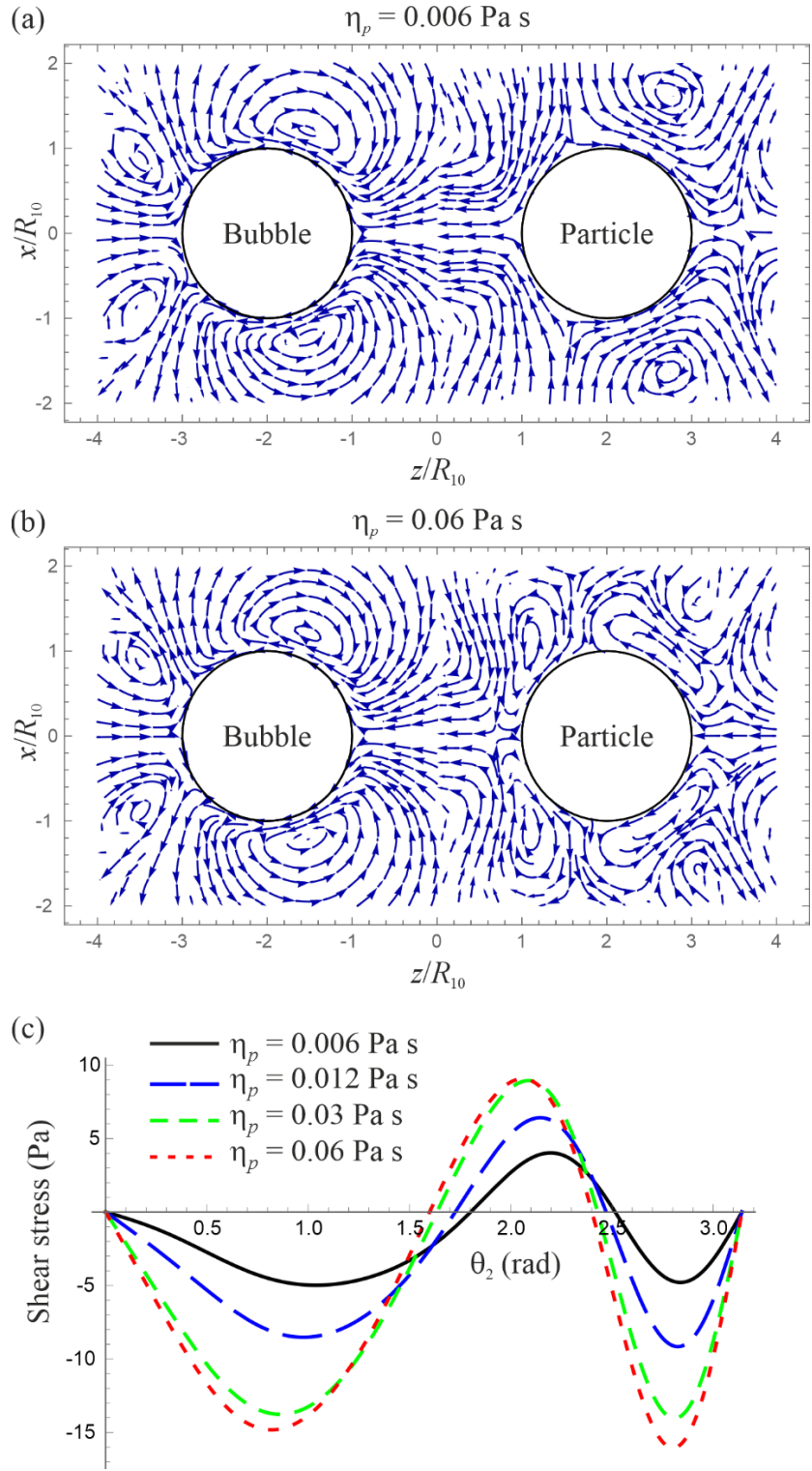

Figure 5. The effect of the particle viscoelasticity. The parameters, except the particle shear viscosity $\eta_p$, are as in figure 4(c). (a) and (b) show acoustic streaming for $\eta_p = 0.006$ Pa s and $\eta_p = 0.06$ Pa s. (c) shows the stress curves for increasing values of $\eta_p$.

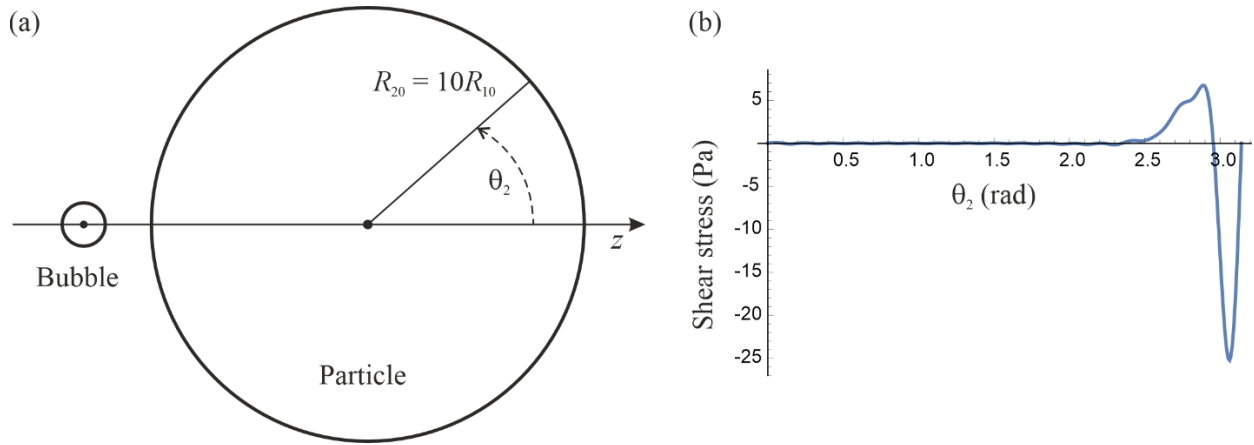

Figure 6. (a) A radially oscillating bubble near a big particle. $R_{10} = 5$ µm, $R_{20} = 50$ µm, $d = 65$ µm, $f = 500$ kHz, $s_0 = 1$ µm. (b) Shear stress produced by the bubble on the particle surface.

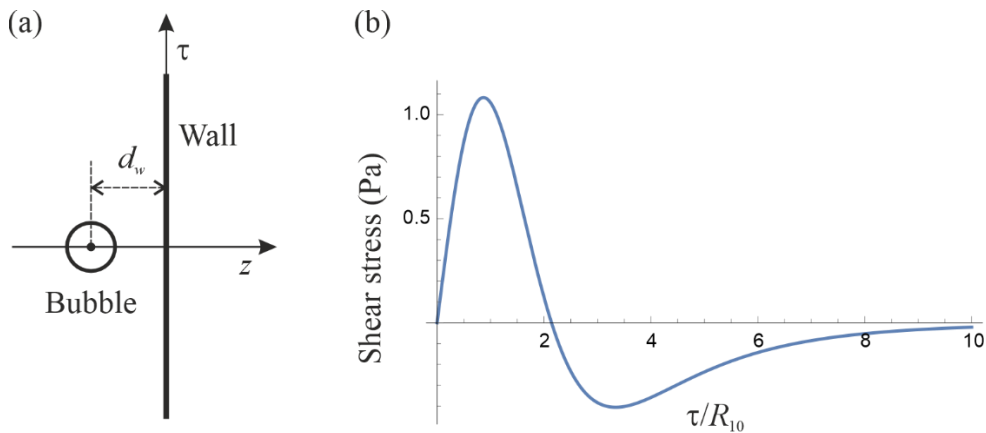

Figure 7. (a) A radially oscillating bubble near a plane rigid wall. $R_{10} = 5$ µm, $d_w = 15$ µm, $f = 500$ kHz, $s_0 = 1$ µm. (b) Shear stress produced by the bubble on the wall, calculated by (3.5).